\documentclass[12pt]{article}

\usepackage{graphicx}

\begin{document}

\title{Towards Robotic Things in Society}
\author{Seng W. Loke\\School of Information Technology, Deakin University, \\Geelong, Australia}

\maketitle

\begin{abstract}
Emerging are so-called smart things embedded with computational, sensing,  networking and actuation capabilities, from smart bins to smart park benches, as well as the proliferation of autonomous vehicles and robots in an increasingly wide range of applications. This is not only an increased in automation affecting and hopefully improving daily life, but also calls for thinking about what a society saturated with such robotic things (i.e., smart things and robots) might look like. This paper discusses five aspects of a vision of Internet connected robotic things (or Internet of Robotic Things (IoRT)) occupying and operating in public spaces, from streets, parks to shopping malls. We discuss, highlighting issues, with the notion of an {\em entourage  of drones and robots accompanying people} in public places, the idea of creating environments or {\em envelopes}  suitable for robot function, the idea of {\em societies of robotic things},   and {\em governance for robotic things} in public spaces. 
\end{abstract}

\section{The Internet of Robotic Things in Public}
The concept of the Internet of Things integrated with robotics has been discussed comprehensively by~\cite{doi:10.1177/1729881418759424} and \cite{7805273}. The combination of connectivity, communication, sensing, and thinking, combined with actuation and the ability to change the physical world in an autonomous way, has profound implications for living in the future. 

As noted in \cite{Loke:2018:WRI:3267305.3274150}, robots are appearing  in public spaces. Some examples of robotic things we would see in public include: delivery robots\footnote{https://www.fastcompany.com/90291820/8-robots-racing-to-win-the-delivery-wars}, robots acting as security guards,\footnote{https://www.knightscope.com/}  self-driving robocarts\footnote{US Patent 20160260161, at http://appft.uspto.gov/netacgi/nph-Parser?Sect1=PTO1\&Sect2=HITOFF\&d=PG01\&p=1\&u=\%2Fnetahtml\%2FPTO\\
\%2Fsrchnum.html\&r=1\&f=G\&l=50\&s1=\%2220160260161\%22.PGNR.\\
\&OS=DN/20160260161\&RS=DN/20160260161}, robots to help with luggages in hotels,\footnote{https://www.telegraph.co.uk/travel/hotels/articles/hotel-robot-room-service/,https://aethon.com/sheraton-hotel-to-use-tug-robots/} automated hotel reception,\footnote{See Japan's robot receptionists at https://asia.nikkei.com/Business/Robot-staff-make-Japan-s-Henn-na-Hotels-quirky-and-efficient and Alibaba's automated hotel at https://kohler.design/flyzoo-hotel/} cleaning robots,\footnote{Vacuum cleaning robots are common and some larger ones used in shopping malls (e.g., in Chadstone, Melbourne, Australia - see https://www.reddit.com/r/melbourne/comments/82nmop/this\_cleaner\_at\_chadstone/) } robots waiting at cafes\footnote{See Shelley robot in a cafe in Ringwood, Melbourne, Australia  at https://www.facebook.com/spacewalkcafe52/posts/shelley-the-robot-waitress-brings-food-to-your-table-at-ringwoods-spacewalk-cafe/2288916167834535/}, healthcare robots~\cite{Riek:2017:HR:3154816.3127874}, and of course, autonomous vehicles. 

Service robots have appeared in many places, such as Pepper\footnote{https://www.softbankrobotics.com/emea/en/robots/pepper}   in Japanese retail shops and in exhibitions as early as 2005~\cite{1546368}, KeJia robot in a shopping mall~\cite{doi:10.1177/1729881417703569}, and a PAL REEM robot trialled in an Australian airport~\cite{Tonkin:2018:DMU:3171221.3171270}.\footnote{See also the project http://being-there.org.uk/}  There is no doubt that the range of use cases and applications will continue to increase. Robots might not need Internet connectivity to function, but often, many commercial service robots connect to a cloud platform and can be easily Internet-enabled, and robots may need to connect to other robots.

Rather than just speaking of robots, we will also use the more general term {\em robotic things}~\cite{8525625,7805273}   in this paper, when we want to refer to both robots in the typical sense, and smart things, i.e., ``everyday objects with autonomous sensing, reasoning and acting capabilities that form part of the Internet of Things - which may not take the typical form of a robot''~\cite{Loke:2017:CMC:3092610}, as in~\cite{Rose14,Kuniavsky:2010:STU:2597837,Gershenfeld:1999:TST:519271,Loke:2018:WRI:3267305.3274150}. We can then include in this term robotic vehicles and drones as well as fixed items such as smart park benches,  smart bookshelves, smart furniture that can change shape automatically, smart homes with  smart walls and shelves, smart street lights,
 smart street signs, or smart sculptures in public, which could move or affect the physical world autonomously in some way, e.g., via robotic arms, and are with or without mobility capabilities.  
 
 Also, what constitutes a robotic thing is also an issue for discussion. For example, a completely automated hotel can be considered a ``smart thing'' but actually comprises a collection of components, from robotic receptionists to robots to handle luggages and room service. The boundaries around a ``smart thing'' and what its constituents are and how smart things could compose to form smart things at larger granularity is a topic we will come back to later in the paper.

This paper discusses five key aspects of a vision of Internet-connected robotic things in society, namely, the notion of an {\em entourage  of drones and robots accompanying people} in public places, the idea of creating environments or {\em envelopes}  suitable for robot function, {\em compositionality} of robotic things,  the idea of {\em societies of robotic things},   and {\em governance for robotic things} in society, especially when they operate in shared public spaces.

\section{Perpectives}

This section discusses five aspects of  the increasing proliferation of robotic things in society, and highlights issues and challenges within.

\subsection{Entourage of Robotic Things and Drones}

An entourage of people normally accompanies royalty or V.I.P.s in their travels.
Robots that help people carry things home have been invented\footnote{For example, https://5elementsrobotics.com/budgee-main/} and also a robot luggage that can follow its owner.\footnote{https://travelmaterobotics.com/}  With the advent of robust human following robots, there could be interesting possibilities. The Temi robot\footnote{https://www.robotemi.com/}  is one example of a personal robot with rather robust   human tracking and following capabilities. Figure~1 illustrates the Temi robot carrying a bag, and the Temi robot as a launchpad for drones.  We may consider such robots in homes and shopping malls, but individuals could take their own robots to shopping and have them follow them around - why one should do that is of course a question, and there are restrictions to the type of ground wheeled robots can travel on. Figure~1 illustrates such an entourage.

While Temi is currently restricted in movement to mainly flat spaces, there are two legged delivery robots being developed,\footnote{https://www.wkbw.com/news/national/ford-is-developing-a-robot-that-would-carry-deliveries-to-your-front-door} and also four-legged robots that work on non-flat terrains such as Laikago,\footnote{http://www.unitree.cc/} Spot,\footnote{https://www.bostondynamics.com/spot} and Anymal,\footnote{https://www.anybotics.com/anymal-legged-robot/}  which can potentially follow its user through all kinds of areas. 

Let us consider a shopping spree scenario. One could imagine an entourage of one or more such robots (and drones), following  the user, and carrying bags of shopping during a shopping spree in the city, or many individuals each with his/her own entourage. A personal drone also follows the shopper to record the experience. One could only drive one car, but multiple robots could follow autonomously. 
A question is then whether this would be socially or publicly acceptable and whether one needs to register such robots (and drones) and purchase permits from the city council in order  to legally take such robots around - similar to how one pays registration and road tax for being able to legally have a vehicle use public roads.  This could also be a deterrent for individuals taking a dozen of such robots in an entourage that could perhaps cause congestion on walkways, or laws might prohibit the size of such an entourage.

Interestingly, some of these robots might mingle with robots in the mall or tour-guide robots in order to enhance the shopping experience for users.  Some of such helper robots might also be a service provided by the shopping centre to ``encourage'' shopping, and take the form of robocarts or robotic trolleys that are offered free-of-charge or or could be rented for a small fee to help shoppers transport goods to their vehicles - other possibilities include automated delivery of bought goods to the shopper's vehicle or even free delivery home.

One could imagine an entourage not only with robocarts but also  robotic prams that automatically fold and unfold and could follow parents around - or auto-wheelchairs for elderly shoppers or elderly parents accompanying shoppers.  Again, shared public spaces can become busy with such robotic things, and would require careful design to comfortably and safely accommodate such robots.

  \begin{figure}
\centering
  
\includegraphics[width=.3\linewidth,height=6.5cm]{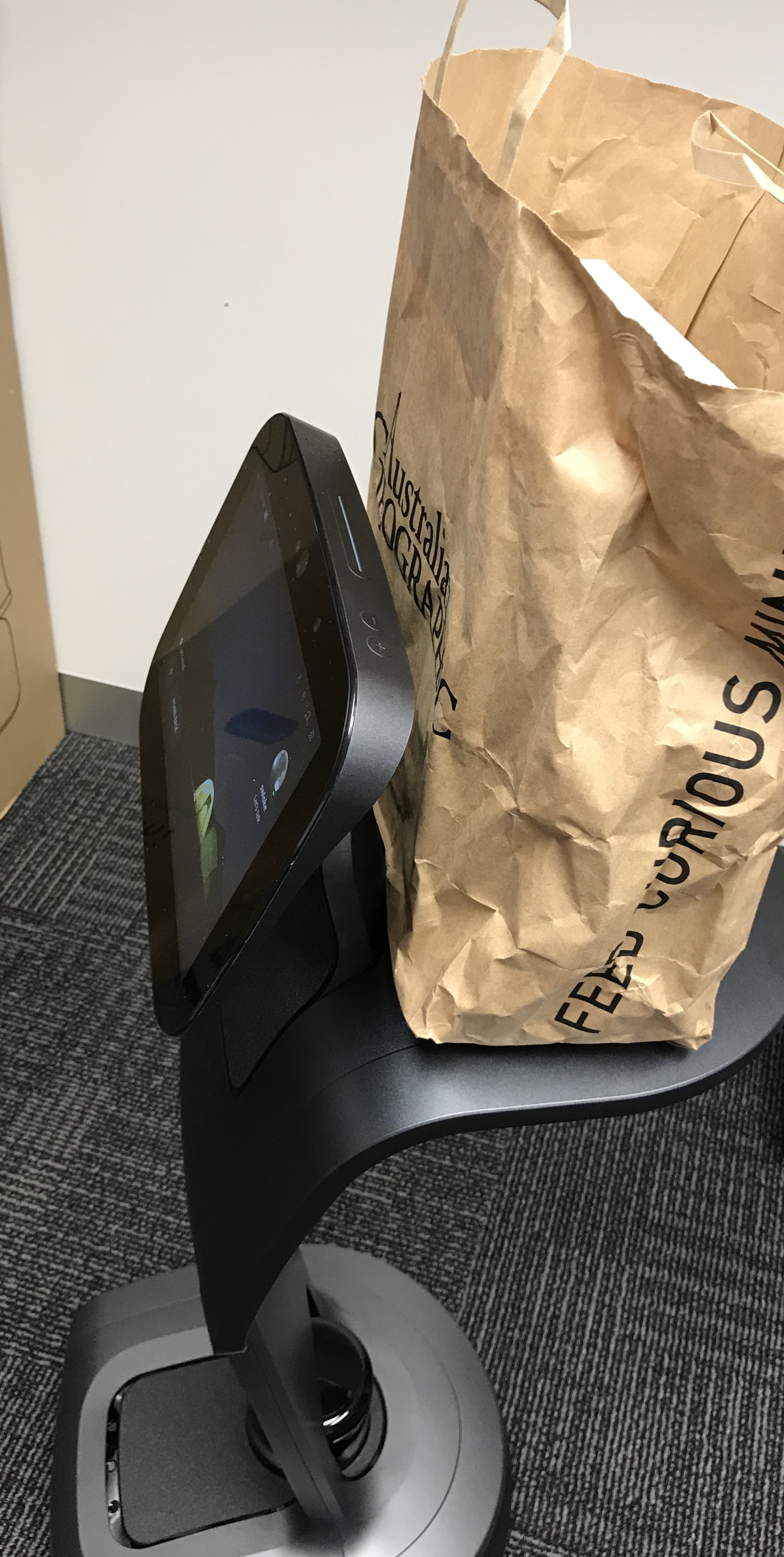}
\includegraphics[width=.3\linewidth,height=6.5cm]{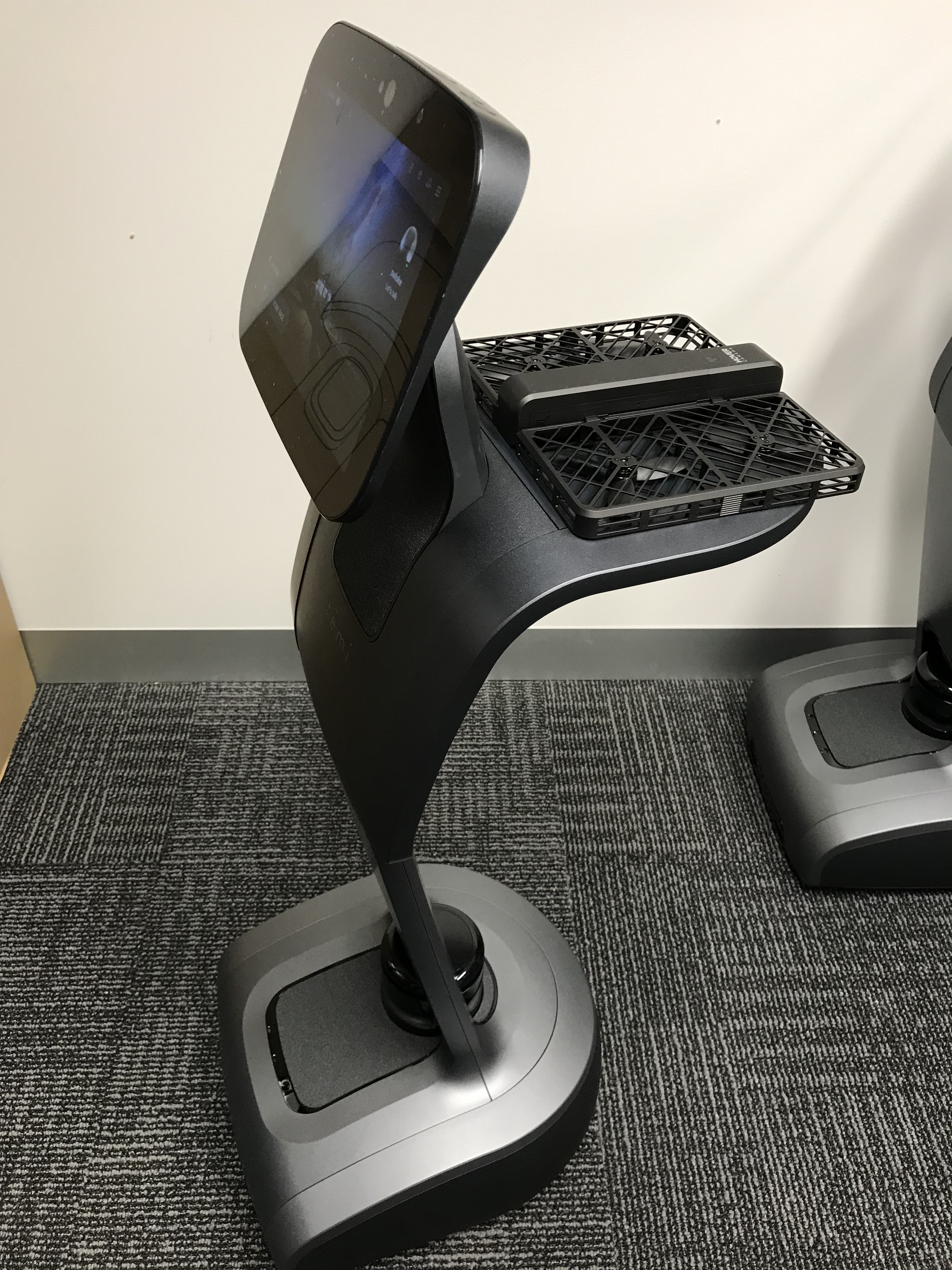}
{\includegraphics[width=.38\linewidth,height=4.5cm]{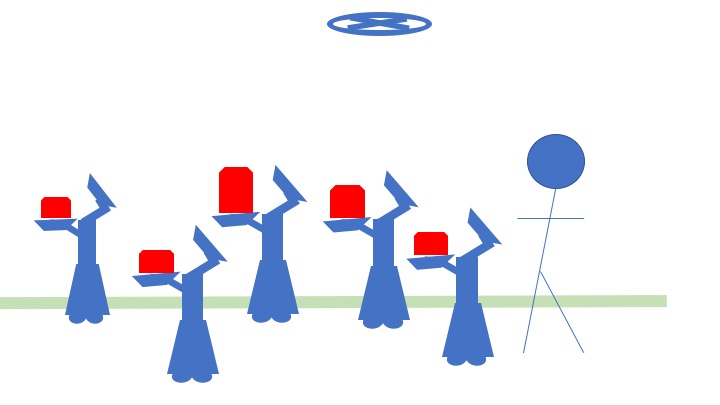}}
     \caption{Temi robot with a bag and as a launchpad for a drone  (Hover - https://gethover.com - in the picture).
     An illustration of an entourage comprising Temi-like shopping robots to help carry goods and a drone (represented by the hovering oval).}
  \label{temi-bag}
 
 

\end{figure}

\subsection{Envelopes and ``Terraforming'' for Robots}

Artificial Intelligence (AI) and robots have their limitations in terms of navigating in physical spaces. The concept of ``enveloping'' was introduced in~\cite{Floridi2019} in regard to providing constraints within which AI can work effectively. From the paper, ``the three-dimensional space that defines the boundaries within which a robot can work successfully is defined as the robot's envelope", and an example is the  dishwasher. Also, one could consider the robots working within a large 
 warehouse. Instrumenting environments or ``terraforming'' for robots can help achieve behaviours otherwise not achievable. Another example is the addition of beacons within an environment to help drones or robots localise accurately and precisely - possible within an environment such as a shopping mall or building. Indeed, robotic things would do better in and be optimised for specialised tasks, and in the short term, not be  versatile enough to work in environments beyond what they have been designed for.

\subsection{Compositionality of Robotic Things}
We imagine the number and range of robotic things at a place growing over time, and it would be useful if the new things could work with the old.
In~\cite{McCullough:2004:DGA:983677}, we have the notion of  piling technologies in a place (e.g., in a room, floor, or building) in a way that they can be organised into a useful whole. Here, we can consider  a system of robotic things which can grow in size and capability over time, and as a system, able to adapt to devices being removed or replaced or added. One can imagine an environment (e.g., an office) where someone buys new robots over time and adds them to the existing collection, and each new robot is able to synergise with the existing collection to form new system capabilities as a whole beyond simple additive capability, basically, devices blending and enhancing each other~\cite{DBLP:conf/aina/SeeraLT07}, so that each existing device is more capable after the addition of the new device.

This is related to the idea of compositionality, where one can also compose different teams of robots from a collection for a certain purpose or function. Imagine a collection of smart picture frames on the wall as in~\cite{DBLP:journals/casa/Loke16}, each of which can sense its neighbouring frame and adjust its image according to what is or is not next to it. Hence, the idea is that one might buy chairs that complement a table, but in the future, one might buy certain robotic things that computationally complement other robotic things - this is just a simple extension of the idea of one buying different components for a computer system (e.g., the CPU, the display, a keyboard, a mouse, a printer and so on).

It would also be useful if the robotic things from the community could be pooled together occasionally to meet community needs - e.g.,  personal robot lawn mowers from multiple houses are pooled together from time to time to mow the grass in a shared park area, or multiple autonomous vehicles of a neighbourhood are volunteered and pooled together in case of an emergency situation.

\subsection{Societies of Robotic Things}

We  consider an analogy with insect societies, which have existed for ages. Insects are one of the most diverse organisms in the world, with with around 900,000 different types of known living insects, making up 80\% of the world's species. They also perform important functions for the environment and ecosystems around the world, from pollinating crops, helping to  create the nutrient-rich layer of soil for plants to cleaning up waste.\footnote{https://entomology.unl.edu/scilit/benefits-insects} To many urban inhabitants, they seem to function largely without heavily encroaching on the lives of people, even if their benefits are great. Within each insect species, there are normally clear roles of different specialised forms of insects, which work together in amazing ways. 

Consider the ant societies, bees, and termites described in~\cite{Camazine:2001:SBS:601161}. Each insect society has members with different roles and swarm behaviours to achieve particular functions.

Analogously, we can consider the notion of {\em robot societies}, each society comprising a collection of robots with different roles and  complementary functions which swarm together to perform larger system functions. Examples include collections of robots for particular tasks:
\begin{itemize}
\item  a collection of robotic things or a robot society for cleaning  a building which might include robots that do vacuuming, that mop the floors, that clean windows and walls and  ceilings, that look for dirty areas, and smart things situated in the building that aid the function of the robots, including sensors and smart cameras, 
\item  a robot society for   rubbish disposal comprising smart bins with wheels/legs that can move into larger bins to transfer waste, and the large bins then move into autonomous vehicles for transport to waste processing sites, with appropriate robots at the waste processing sites to sort and process the waste and so on; this robot society might work with the robot society for cleaning at certain points,
\item autonomous restaurants, with robots for cooking, taking orders, serving food, complemented by a robot society that cleans the restaurant, a robot society that grows the food (e.g., comprising farm robots and other robotic things), and a robot society responsible for transporting food from the farms to the restaurants,
\item a robot society comprising autonomous taxis, autonomous trains, autonomous wheelchairs which handle transportation needs for an area,
\item a robot society for an aged care facility, comprising not only robots to clean the building, but also to take care (to an extent) of individuals with needs,
\item a robot society for the library, from cleaning the entire library building to sorting out books, retrieving books, orientating users, and handling queries.
\end{itemize}
There are many more possible robot societies one could think of, which might be designed to work without encroaching into the lives of people, somewhat ``in the background''. Humans might also work with such robot societies at certain points, where human attention is required. It is interesting to consider the size and granularity of such robot societies, and perhaps a hierarchical structure, the robot society for cleaning a building might comprise collections (further specialised robot societies) of window cleaning robots, collections of robots for wall cleaning, for floor cleaning and so on.

\subsection{Layered Models of Behaviour for Robotic Things}
\begin{figure}[h!]
\begin{center}
\includegraphics[width=10cm]{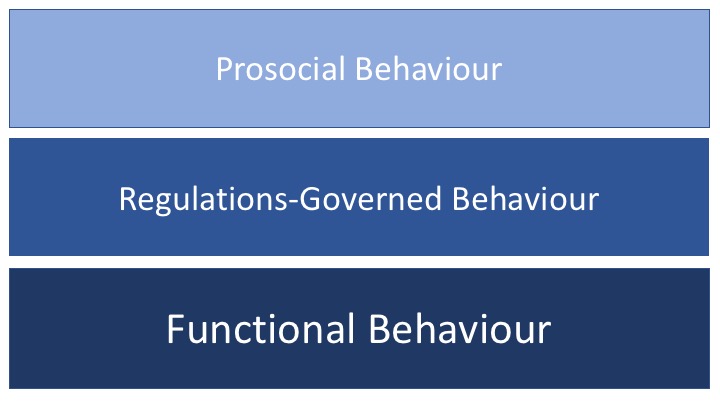}
\end{center}
\caption{ Layers of Behaviour for  Robotic Things in Public}\label{layers}
\end{figure}

 We can categorize the range of required behaviours of robotic things in public into layers, as first introduced in~\cite{Loke:2018:WRI:3267305.3274150}, as shown in  Figure~\ref{layers}.
 The bottom layer refers to the correct functioning of the robotic thing according to its {\em purpose}. For example, a walking path cleaning robot should indeed clean the path properly.
 The second layer refers to {\em regulations-governed behaviour}, which might constrain operations. For example, the town council's path cleaning robot does not  clean paths in privately own properties, and must stay within public paths.
 The top layer refers to {\em good or pro-social behaviour}, where, for instance,  the path cleaning robot ``politely'' stops to give way to another robot or human walking past, not just cleaning paths well or following local regulations. Imagine a personal entourage of robots moving along with  commercial delivery robots along walkways, in a way that is respectful and considerate.

Think of a talking smart wall which lights up and announces advertisements as people pass near it - the wall will need to function correctly providing the right ads as programmed, but also will need to behave in ways allowed by local regulations, e.g., in terms of what form  in which to make announcements (e.g., multi-lingual) and what frequency. When the wall is in operation, it should ideally be able to also lower its volume sensing conversations around it.

We might expect such behaviours also of autonomous vehicles, which need to perform their function of taking passengers to the right places, but also in a way which regulations are followed, and in a way that is safe and respecting of not only its passengers but other drivers (e.g., not driving in an unsafe way, even if within legal limits).

\section{Conclusion}
We have sought to answer the question of the implications of having robotic things in public, in terms of five aspects of a vision of Internet-connected robotic things in public: (i) the idea of an entourage of robotic things surrounding users, (ii) the concept of envelopes and creating environments for the operation of robotic things, (iii) the idea of compositional robotic things at a place, (iv) the idea of societies of robotic things, and (v) layered models for safe and prosocial robotic things in public. Each aspect has its own challenges, e.g., rules to determine what might be an acceptable entourage, the engineering challenges in designing and developing such envelopes and environments for robotic things to thrive and synergize, the design and creation of robotic societies, and governance, given that such robotic things will operate in shared spaces. There are also questions regarding how to  design future cities, or redesign neighbourhoods, in order to accommodate the ``infrastructure'' of robot societies that manage and function in the city.

There are many other aspects of the Internet of robotic things discussed elsewhere, including ethical concerns and security and privacy issues (e.g., see~\cite{ALLHOFF201855}). The paper has not sought to provide a complete review of all possible perspectives or solutions, but to identify issues and initiate discussion on socio-technical issues (which perhaps has not been given adequate attention elsewhere) of a society potentially saturated with robotic things. 



\bibliographystyle{plain} 

\bibliography{mybib,robots}


\end{document}